\documentclass[cits,a4paper]{Pos}

\usepackage{amsmath,amssymb}
\usepackage{subfigure}
\usepackage{graphicx}

\newcommand{\mps}{m_\mathrm{PS}}
\newcommand{\fps}{f_\mathrm{PS}}

\newcommand{\nft}{N_\mathrm{f}=2}
\newcommand{\nftp}{N_\mathrm{f}=2+1+1}
\newcommand{\preprint}{\newline%
  \begin{picture}(0,0)
  \put(370,100){\rm\small DESY 09-226}
  \end{picture}}

\graphicspath{{plots/}}

\title{Scaling and $\chi$PT Description of Pions
              from $\nft$ twisted mass QCD
              \preprint
}

\ShortTitle{Scaling and $\chi$PT Description of Pions from $\nft$
  tmQCD}

\author{Petros Dimopoulos, Roberto Frezzotti\\
  Dipartimento di Fisica, Universit\`a di Roma ``Tor Vergata''\\
  Via della Ricerca Scientifica 1, 00133 Rome, Italy\\
  E-mail: \email{\{dimopoulos,frezzotti\}@roma2.infn.it}}

\author{\speaker{Gregorio Herdoiza}, Karl Jansen\\
  NIC, DESY\\
  Platanenallee 6, 15738 Zeuthen, Germany\\
  E-mail: \email{\{Gregorio.Herdoiza,Karl.Jansen\}@desy.de}}

\author{Chris Michael\\
  Theoretical Physics Division, Department of Mathematical Sciences,
  University of Liverpool\\
  Liverpool L69 3BX, UK\\
  E-mail: \email{c.michael@liverpool.ac.uk}}

\author{Carsten Urbach\\
  Helmholtz-Institut f{\"u}r Strahlen- und Kernphysik
  (Theorie) and\\
  Bethe Center for Theoretical Physics, Universit{\"a}t
  Bonn\\
  53115 Bonn, Germany\\
  E-mail: \email{urbach@hiskp.uni-bonn.de}}

\author{for the ETM Collaboration}

\abstract{We study light-quark observables by means of dynamical
  lattice QCD simulations using two flavours of twisted mass fermions
  at maximal twist. We employ chiral perturbation theory to describe
  our data for the pion mass and decay constant. In this way, we
  extract precise determinations for the low-energy constants of the
  effective theory as well as for the light-quark mass and the chiral
  condensate.}

\FullConference{The XXVII International Symposium on Lattice Field
  Theory - LAT2009\\ July 26-31 2009\\ Peking University, Beijing,
  China}

\begin{document}

\section{Introduction}

In recent years, the non-perturbative description of QCD on the
lattice has made a significant breakthrough in tackling the systematic
effects present in the determination of several important physical
quantities, opening the way for a direct connection to experiments
(see {\it e.g.} \cite{Jansen:2008vs, Scholz:2009yz} for recent
reviews). Simulations containing the dynamics of the light-quark
flavours in the sea, as well as those due to the strange quark and
recently also to the charm, using pseudoscalar masses below
$300$\,MeV, lattice extents $L \geq 2$\,fm and lattice spacings
smaller than $0.1$\,fm are presently being performed by several
lattice groups. Such simulations will eventually allow for an
extrapolation of the lattice data to the continuum limit and to the
physical point while keeping also the finite volume effects under
control.

The European Twisted Mass collaboration (ETMC) has carried out large
scale simulations with $\nft$ flavours of mass degenerate quarks using
Wilson twisted mass fermions at maximal twist. Four values of the
lattice spacing ranging from $0.1$\,fm down to $0.051$\,fm,
pseudoscalar masses between $280$ and $650$\,MeV as well as several
lattice sizes ($2.0 - 2.5$\,fm) are used to address the systematic
effects.

The physics of the light pseudoscalar meson is in a suitable sector
for investigating the systematic effects arising from the continuum,
infinite volume and chiral extrapolations of lattice data, since the
pion mass and decay constant can be measured with high statistical
accuracy in lattice simulations.  Moreover, chiral perturbation theory
($\chi$PT) is best understood for those two quantities. As a result of
this study one can extract several important quantities, such as the
$u,d$ quark masses, the chiral condensate or the low-energy constants
of $\chi$PT. First results for the pseudoscalar mass $\mps$ and decay
constant $\fps$ from these $\nft$ simulations have been presented in
Refs.~\cite{Boucaud:2007uk, Urbach:2007rt, Dimopoulos:2007qy,
  Boucaud:2008xu, Dimopoulos:2008sy, Dimopoulos:2009xx, scalingnf2}.

ETMC is currently generating $N_{\rm f}=2+1+1$ ensembles including in
the sea, in addition to the mass degenerate light $u,d$ quark
flavours, also the heavier strange and charm degrees of freedom.
First results for the pseudoscalar mass and decay constant from this
novel setup were presented in~\cite{Baron:2008xa,Baron:2009zq}.

In the following we will mainly focus on the results from the $\nft$
data for $\mps$ and $\fps$.

\section{Lattice Action and Simulation Setup}

In the gauge sector we employ the tree-level Symanzik improved gauge
action (tlSym)~\cite{Weisz:1982zw}. The fermionic action for two
flavours of maximally twisted, mass degenerate
quarks~\cite{Frezzotti:2000nk,Frezzotti:2003ni} in the so-called
twisted basis (where the quark field doublets are denoted by $\chi$
and $\bar\chi$) reads
\begin{equation}
  \label{eq:sf}
  S_\mathrm{tm}\ =\ a^4\sum_x\left\{ \bar\chi(x)\left[ D[U] + m_0 +
      i\mu_q\gamma_5\tau^3\right]\chi(x)\right\}\, ,
\end{equation}
where $m_0$ is the untwisted bare quark mass tuned to its critical
value $m_\mathrm{crit}$, $\mu_q$ is the bare twisted quark mass,
$\tau^3$ is the third Pauli matrix acting in flavour space and $D[U]$ 
is the Wilson-Dirac operator.

At maximal twist, i.e.~$m_0=m_\mathrm{crit}$, physical observables are
automatically O$(a)$ improved without the need to determine any action
or operator specific improvement coefficients~\cite{Frezzotti:2003ni}
(for a review see Ref.~\cite{Shindler:2007vp}). With this being the
main advantage\,\footnote{Other properties being, {\it e.g.}, that the
  quark mass renormalises only multiplicatively and that the
  determination of the pseudoscalar decay constant does not require a
  renormalisation factor.}, one drawback of maximally twisted mass
fermions is that parity and flavour symmetry are broken explicitly at
finite values of the lattice spacing, which amounts to O$(a^2)$
effects in physical observables.

For details on the setup, tuning to maximal twist and the analysis
methods we refer to Refs.~\cite{Boucaud:2007uk, Urbach:2007rt,
  Boucaud:2008xu}. Recent results for light quark masses, meson decay
constants, the pion form factor, $\pi$-$\pi$ scattering, the light
baryon spectrum, the $\eta'$ meson and the $\omega-\rho$ mesons mass
difference are available in Refs.~\cite{Blossier:2007vv,
  Frezzotti:2008dr, Feng:2009ij, Alexandrou:2008tn, Jansen:2008wv,
  McNeile:2009mx, Blossier:2009bx}.

Flavour breaking effects have been investigated for several
quantities~\cite{Boucaud:2007uk, Urbach:2007rt,Boucaud:2008xu,
  Dimopoulos:2008sy, Alexandrou:2008tn}. With the exception of the
splitting between the charged and neutral pion masses, other possible
splittings in the unitary observables so far investigated are found to
be compatible with zero. These results are in agreement with a
theoretical investigation using the Symanzik effective
Lagrangian~\cite{Frezzotti:2007qv, Dimopoulos:2009qv}.

A list of the $\nft$ ensembles generated by ETMC can be found in
table~\ref{tab:setup}.

\begin{table}[t!]
  \centering
  \begin{tabular*}{0.85\textwidth}{@{\extracolsep{\fill}}ccrcccr}
    \hline\hline
    $\Bigl.\Bigr.$ Ensemble &$\beta$ & $a$~[fm] &$V/a^4$  
    & $\mps L$ & $a\mu_q$
    & $\mps$~[MeV]  \\ \hline\hline
    $D_1$ & $4.20$ & $ 0.051$ & $48^3 \cdot 96$ & $3.6$ & $0.0020$ & $280$ \\
    $D_2$ & & & $32^3 \cdot 64$ & $4.2$ & $0.0065$ & $510$  \\
    \hline
    $C_1$ & $4.05$ & $ 0.063$ & $32^3 \cdot 64$ & $3.3$ & $0.0030$ & $320$ \\
    $C_2$ & & & & $4.6$  & $0.0060$ & $450$ \\
    $C_3$ & & & & $5.3$ & $0.0080$ & $520$ \\
    $C_4$ & & & & $6.5$ & $0.0120$ & $630$ \\
    $C_5$ & & & $24^3 \cdot 48$& $3.5$ & $0.0060$ & $450$ \\
    $C_6$ & & & $20^3 \cdot 48$ & $3.0$ & $0.0060$ & $450$ \\
    \hline
    $B_1$ & $3.90$ & $ 0.079$ & $24^3 \cdot 48$ & $3.3$ & $0.0040$ & $330$ \\
    $B_2$ & & & & $4.0$ & $0.0064$ & $420$ \\
    $B_3$ & & & & $4.7$ & $0.0085$ & $480$ \\
    $B_4$ & & & & $5.0$ & $0.0100$ & $520$ \\
    $B_5$ & & & & $6.2$ & $0.0150$ & $640$ \\
    $B_6$ & & & $32^3 \cdot 64$ & $4.3$ & $0.0040$ & $330$ \\
    $B_7$ & & & & $3.7$ & $0.0030$ & $290$ \\
    \hline
    $A_2$ & $3.80$ & $ 0.100$ & $24^3 \cdot 48$ & $5.0$ & $0.0080$ & $ 410$ \\
    $A_3$ & & & & $5.8$ & $0.0110$ & $ 480$ \\
    $A_4$ & & & & $7.1$  & $0.0165$ & $ 580$ \\
    \hline\hline
  \end{tabular*}
  \caption{Ensembles with $\nft$ dynamical flavours produced by the
    ETM collaboration.  We give the ensemble name, the values of the
    inverse bare coupling $\beta=6/g_0^2$, an approximate value of the
    lattice spacing $a$, the lattice volume $V=L^3 \cdot T$ in lattice
    units, the approximate value of $\mps L$, the bare quark mass
    $\mu_q$ in lattice units and an approximate value of the light
    pseudoscalar mass $\mps$.}
  \label{tab:setup}
\end{table}

\section{Scaling to the Continuum Limit}

Here we analyse the scaling to the continuum limit of the pseudoscalar
meson decay constant $\fps$ at fixed reference values of the
pseudoscalar meson mass $\mps$ and of the lattice size $L$ (we refer
to~\cite{Urbach:2007rt, Dimopoulos:2007qy, scalingnf2} for
details). The purpose of this scaling test is to verify that
discretisation effects are indeed of $O(a^2)$ as expected for twisted
mass fermions at maximal twist.

In order to compare results coming from different values of the
lattice spacing it is convenient to measure on the lattice the
hadronic scale $r_0$ \cite{Sommer:1993ce}. It is defined via the force
between static quarks at intermediate distance and can be measured to
high accuracy in lattice QCD simulations. For details on how we
measure $r_0/a$ and on how its chiral extrapolation is performed, we
refer to Ref.~\cite{Boucaud:2008xu, scalingnf2}.

\begin{figure}[t]
  \centering
  \subfigure[\label{fig:fpsscaling}]%
  {\includegraphics[width=0.45\linewidth]{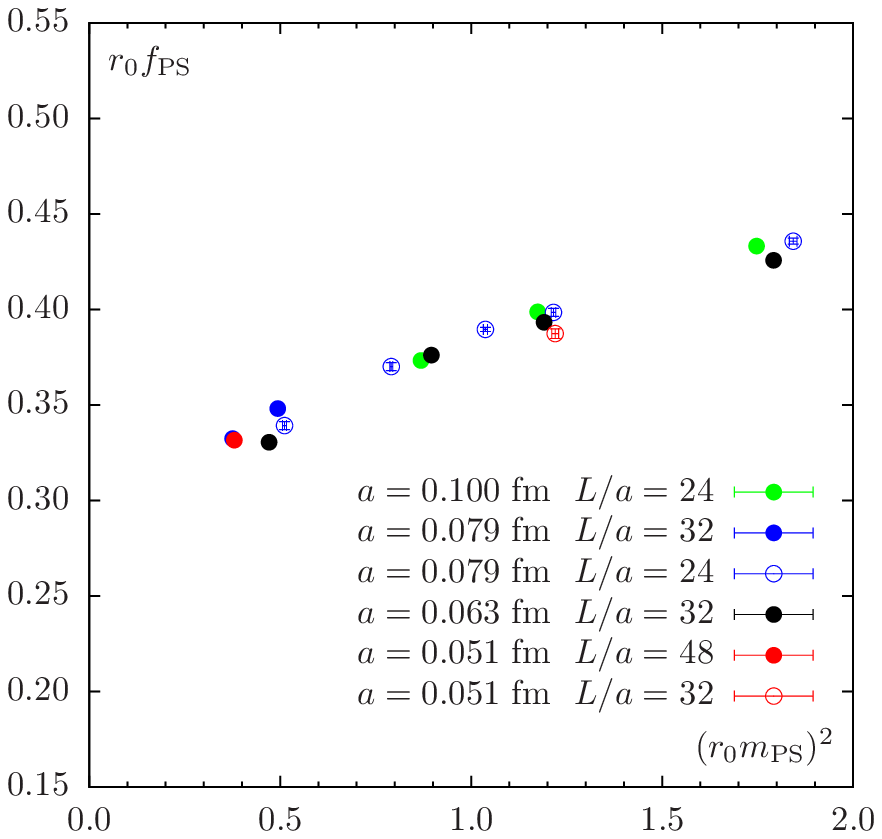}}
  \qquad
  \subfigure[\label{fig:scaling}]%
  {\includegraphics[width=0.455\linewidth]{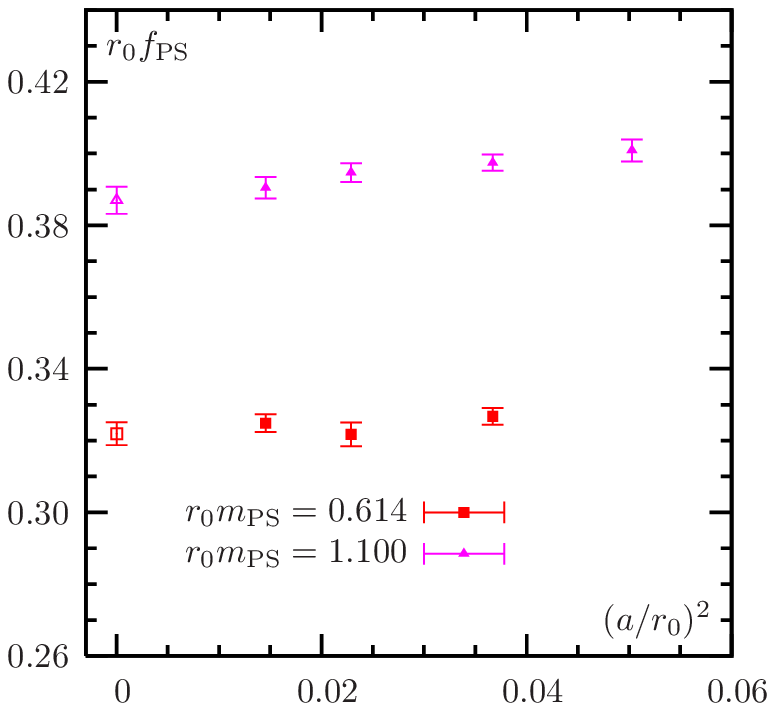}}
  \caption{(a) $r_0\fps$ as a function of $(r_0\mps)^2$. (b) Continuum
    limit scaling:~$r_0\fps$ as a function of $(a/r_0)^2$ at two fixed
    values of $r_0\mps$ and at fixed volume.}
  \label{fig:latart}
\end{figure}

In figure~\ref{fig:fpsscaling} we plot the raw data for $r_0\fps$ as a
function of $(r_0\mps)^2$. The vicinity of points coming from
different lattice spacings along a common curve is an evidence that
lattice artifacts are small for these quantities. This is indeed
confirmed in figure~\ref{fig:scaling} where the continuum scaling of
$r_0\fps$ is illustrated:~the very mild slope of the lattice data
shows that the expected O$(a^2)$ scaling violations are small. The
result of a linear extrapolation in $(a/r_0)^2$ to the continuum limit
is also shown.

Analogous continuum limit scaling studies have been performed for the
charged pion mass~\cite{Dimopoulos:2007qy, scalingnf2} and the nucleon
mass~\cite{Alexandrou:2008tn, Alexandrou:2009qu}, showing also in these
cases signs of only small scaling violations.

\section{$\chi$PT Description of Lattice Data}

The chiral extrapolation of lattice data down to the physical point is
currently one of the main sources of systematic uncertainties in the
lattice results. The possibility to rely on an effective theory such
as $\chi$PT to guide this extrapolation is therefore of great
importance to quote accurate results from lattice simulations. On the
other hand, while smaller quark masses are becoming accessible in
numerical simulations, the possibility to perform a quantitative test
of the effective theory as well as to determine the low energy
parameters of its Lagrangian becomes more and more realistic.

We now proceed by presenting the results of a combined continuum,
chiral and infinite volume extrapolation of $\mps$ and $\fps$ for two
values of the lattice spacing (corresponding to $a \approx 0.063$\,fm
and $a\approx 0.079$\,fm). We use the chirally extrapolated values of
$r_0/a$ to relate data from the two lattice spacings and a
non-perturbative determination of the renormalisation factor
$Z_\mathrm{P}$~\cite{Dimopoulos:2007fn} in order to perform the fit in
terms of renormalised quark masses. This analysis closely follows
those presented in Refs.~\cite{Urbach:2007rt, Dimopoulos:2007qy,
  Dimopoulos:2008sy, Dimopoulos:2009xx, scalingnf2} to which we refer
for more details.

We perform combined fits to our data for $\fps$, $\mps$, $r_0/a$ and
$Z_\mathrm{P}$ at the two values of the lattice spacing with the formulae:
\begin{equation}
  \label{eq:fmps}
  \begin{split}
    r_0\fps &= r_0
    f_0\Bigl[1-2\xi\log\left(\frac{\chi_\mu}{\Lambda_4^2}\right) +
    T_f^\mathrm{NNLO} + D_{f_\mathrm{PS}}(a/r_0)^2\Bigr]\
    K_f^\mathrm{CDH}(L)\, , \\
    (r_0 \mps)^2 &= \chi_\mu r_0^2\Bigl[
    1+\xi\log\left(\frac{\chi_\mu}{\Lambda_3^2}\right)+
    T_m^\mathrm{NNLO} + D_{m_\mathrm{PS}}(a/r_0)^2\Bigr]\
    K_m^\mathrm{CDH}(L)^2\, ,\\
  \end{split}
\end{equation}
with $ \xi \equiv 2B_0\mu_R/(4\pi f_0)^2\ ,\,\chi_\mu\equiv
2B_0\mu_R\ ,\, \mu_R \equiv \mu_q/Z_\mathrm{P},\, f_0\equiv\sqrt{2}
F_0$.  $T_{m,f}^\mathrm{NNLO}$ denote the continuum NNLO terms of the
chiral expansion~\cite{Leutwyler:2000hx}, which depend on
$\Lambda_{1-4}$ and $k_M$ and $k_F$. The finite size corrections
factors $K_{m,f}^\mathrm{CDH}(L)$ refer to a continuum $\chi$PT
description using the resummed L{\"u}scher formula, that we denote as
CDH ~\cite{Colangelo:2005gd}.\,\footnote{For a more detailed
  description of finite size effects in our data for $\mps$ and
  $\fps$, we refer to Refs.~\cite{Urbach:2007rt, scalingnf2}.}

Based on the form of the Symanzik expansion in the small quark mass
region, we parametrise in eq.~(\ref{eq:fmps}) the leading cut-off
effects by the two coefficients $D_{f_{\rm PS},m_{\rm PS}}$. Setting
$D_{f_{\rm PS},m_{\rm PS}}=0$ is equivalent to perform a constant
continuum extrapolation. Similarly, setting $T_{m,f}^\mathrm{NNLO}=0$
corresponds to fit to NLO $\chi$PT.

From the fit parameters related to the quark mass dependence
predicted by $\chi$PT (in particular from $\Lambda_{3,4}$, $B_0$ and
$f_0$) the low energy constants
$\bar\ell_{3,4}$ and the chiral
condensate $\Sigma$ can be determined, using the following expressions:
\begin{equation}
  \label{eq:lec}
  \begin{split}
  \bar{\ell}_i =& \log\left(\frac{\Lambda_i^2}{(m_\pi^\pm)^2}\right)\, , \\
  \Sigma =& -\frac{B_0 f_0^2}{2} \,.\\
  \end{split}
\end{equation}

By including or excluding data points for the heavier quark masses, it
is in principle possible to explore the regime of masses in which NLO
and/or NNLO SU$(2)~\chi$PT applies. We have actually generalised this
procedure in order to estimate all the dominant sources of systematic
uncertainties that can be addressed from our setup, which include,
discretisation effects, the order at which we work in $\chi$PT or
finite size effects. The idea is to use different fit ans{\"a}tze (see
below) on a given data-set and to repeat this same procedure over
different data-sets: by weighting all these fits by their confidence
level we construct their distribution and estimate the systematic
error from the associated 68\% confidence interval.

The fit ans{\"a}tze we consider are:
\begin{itemize}
\item Fit A: NLO continuum $\chi$PT, $T_{m,f}^\mathrm{NNLO}\equiv0$,
  $D_{\mps,\fps}\equiv0$, priors for $r_0\Lambda_{1,2}$

\item Fit B: NLO continuum $\chi$PT, $T_{m,f}^\mathrm{NNLO}\equiv0$,
  $D_{\mps,\fps}$ fitted, priors for $r_0\Lambda_{1,2}$

\item Fit C: NNLO continuum $\chi$PT, $D_{\mps,\fps}\equiv0$,
  priors for $r_0\Lambda_{1,2}$ and $k_{M,F}$

\item Fit D: NNLO continuum $\chi$PT, $D_{\mps,\fps}$
  fitted, priors for $r_0\Lambda_{1,2}$ and $k_{M,F}$
\end{itemize}

The choice of the different data-sets (each of them including data for
different lattice spacings, quark masses and physical volumes) is made
in order to quantify how the quality of the fit is modified when
including/excluding data from {\it e.g.}, a given mass region or with
a given volume. The data-sets considered in the fits are listed in
Ref.~\cite{scalingnf2}.

The parameters $k_{M,F}$ and $r_0\Lambda_{1,2}$ appearing at NNLO (and
in the latter case, also at higher order in the CDH expressions) need
to be fitted with some additional input information
(priors). Additional lattice data would be needed to allow these
parameters to remain free in the fit. The parameters
$r_0\Lambda_{1,2}$ use priors from the estimates for $\bar\ell_{1,2}$
in Ref.~\cite{Colangelo:2005gd}, while very mild priors are used for
$k_{M,F}=0\pm 10$. For a detailed description of the use of priors
and, in general, of the statistical analysis, we refer to
Ref.~\cite{scalingnf2}.

The results of the combined fits to data from the two lattice spacings
$a~\approx~0.063$\,fm and $a\approx~0.079$\,fm are given in
table~\ref{tab:results}. The errors are statistical and systematical,
added in quadrature. For a review on the determination of the
low-energy constants and of the light-quark mass we refer
to~\cite{Aokicd09, Leutwylercd09, Scholz:2009yz}. Recent determinations
from ETMC of the pseudoscalar decay constant and chiral condensate in
the $\epsilon$-regime were presented in~\cite{Jansen:2009tt}.

\begin{table}[t!]
  \centering
  \begin{tabular*}{.5\linewidth}{@{\extracolsep{\fill}}lr}
    \hline\hline
    $m_{u,d}\ [\mathrm{MeV}]$      & $3.54(26)$   \\
    $\bar\ell_3$                   & $3.50(31)$   \\
    $\bar\ell_4$                   & $4.66(33)$   \\
    $f_0\ [\mathrm{MeV}]$          & $121.5(1.1)$ \\
    $f_\pi/f_0$                    & $1.0755(94)$ \\
    $B_0\ [\mathrm{MeV}]$          & $2638(200)$  \\
    $|\Sigma|^{1/3}\ [\mathrm{MeV}]$ & $270(7)$ \\
    \hline\hline
   \end{tabular*}
  \caption{Summary of fit results. The errors are statistical and
    systematical, added in quadrature. $B_0$, $\Sigma$ and $m_{u,d}$
    are (non-perturbatively) renormalised in the
    $\overline{\mathrm{MS}}$ scheme at the scale $\mu =
    2\ \mathrm{GeV}$.}
  \label{tab:results}
\end{table}

As a further check, we have performed additional fits including either
the finer lattice spacing ($a\approx 0.051$\,fm) or the coarser one
($a\approx 0.100$\,fm) finding total compatibility with the results of
table~\ref{tab:results}. 

\section{Discussion and Conclusion}

Here we collect a short list of observations coming from a set of
$\chi$PT fits. A complete description of these fits was
presented in Ref.~\cite{scalingnf2}.

We observe that including in the fits pseudoscalar masses $\mps >
520\ \mathrm{MeV}$ decreases significantly the quality of the NLO fits
($\chi^2/\mathrm{dof} \gg 1$). This indicates that the applicability
of NLO $\chi$PT in that regime of masses is disfavoured.

On the contrary, extending the fit-range to a value of $\mps\sim
280\ \mathrm{MeV}$ preserves the good quality of the fit and gives
compatible values for the fit parameters. This result makes us
confident that the extrapolation to the physical point is trustworthy.

Including lattice artifacts in the fits gives results which are
compatible to those where $D_{\mps,\fps}$ is set to zero but with a
somehow better $\chi^2/\mathrm{dof}$. We observe that the values of
the fitting parameters $D_{\mps,\fps}$ are compatible with zero within
two standard deviations. This is in line with the small discretisation
effects observed in the scaling test.

The inclusion of NNLO terms produces similar results to the NLO fits
in the quark mass region corresponding to $\mps \in
[280,520]$\,MeV. When fitting data only in this mass region ({\em
  i.e.}  when excluding from the fit the heavier masses at $\mps \sim
650$\,MeV), we observe that the fit curve at NLO lies closer to those
data points (heavier masses) than the NNLO one. On the other hand,
when including the heavier masses, the NNLO fit is able describe these
data points but the quality of the fit is somehow reduced with respect
to the NLO fit. To improve the sensitivity of our lattice data to
$\chi$PT at NNLO, additional data points would be needed.

We have presented determinations of $\fps$ and $\mps$ and of their
continuum, infinite volume and chiral extrapolations. As a result, we
obtain accurate determinations of the $u,d$ quark mass, the chiral
condensate as well as of low-energy constants of the effective theory,
including an exhaustive estimate for the systematic uncertainties. 

The only systematic effect which cannot be addressed by this study
corresponds to effect of the strange and the charm quarks in the
sea. We are currently in the process of extending this analysis to the
$\nftp$ ensembles which are being generated by
ETMC~\cite{Baron:2008xa,Baron:2009zq}.

We thank the members of ETMC for the most enjoyable collaboration. 

\bibliographystyle{h-physrev5}
\bibliography{bibliography}

\end{document}